\documentclass[sigconf]{acmart}

\setcopyright{cc}
\setcctype{by}
\copyrightyear{2026}
\acmYear{2026}
\acmConference[PEARC '26]{Practice and Experience in Advanced Research Computing}{July 26--30, 2026}{Minneapolis, MN, USA}
\acmBooktitle{Practice and Experience in Advanced Research Computing (PEARC '26), July 26--30, 2026, Minneapolis, MN, USA}
\acmISBN{979-8-4007-2377-3/2026/07}
\acmDOI{10.1145/3785462.3815847}

\begin{document}


\title{STREAM: Multi-Tier LLM Inference Middleware with Dual-Channel HPC Token Streaming}

\author{Anas Nassar}
\authornote{Corresponding author.}
\orcid{0009-0008-4225-5745}
\affiliation{
  \department{Advanced Cyberinfrastructure for Education and Research (ACER)}
  \institution{University of Illinois Chicago}
  \city{Chicago}
  \state{Illinois}
  \country{USA}
}
\email{nassar@uic.edu}

\author{Steve Mohr}
\orcid{0009-0009-0455-8216}
\affiliation{
  \department{Advanced Cyberinfrastructure for Education and Research (ACER)}
  \institution{University of Illinois Chicago}
  \city{Chicago}
  \state{Illinois}
  \country{USA}
}
\email{smohr@uic.edu}

\author{Leonard Apanasevich}
\orcid{0000-0002-5685-5871}
\affiliation{
  \department{Advanced Cyberinfrastructure for Education and Research (ACER)}
  \institution{University of Illinois Chicago}
  \city{Chicago}
  \state{Illinois}
  \country{USA}
}
\email{apana@uic.edu}

\author{Himanshu Sharma}
\orcid{0000-0002-7498-8053}
\affiliation{
  \department{Advanced Cyberinfrastructure for Education and Research (ACER)}
  \institution{University of Illinois Chicago}
  \city{Chicago}
  \state{Illinois}
  \country{USA}
}
\email{himanshu@uic.edu}


\begin{abstract}
Researchers and practitioners working with large language models face a fragmented landscape: local models are free and private but hardware limits the model size and context windows a researcher can use; institutional HPC centers offer powerful GPU resources at no marginal cost and keep data within institutional boundaries, but operate behind firewalls and are designed for batch jobs rather than interactive use; commercial cloud APIs provide frontier-model quality on demand but impose significant cost and data retention policies unsuitable for sensitive research data. No existing system unifies all three. STREAM (Smart Tiered Routing Engine for AI Models) addresses this gap with four contributions: (1)~a three-tier routing architecture combining local, HPC, and cloud inference with a local LLM-based complexity judge; (2)~a dual-channel HPC streaming architecture that separates the Globus Compute control plane (authentication and job dispatch) from a WebSocket relay data plane (token delivery), enabling sub-second TTFT (0.54~s median, 21.1$\times$ over batch mode's 11.40~s) through institutional firewalls without VPN or firewall rule changes, with end-to-end AES-256-GCM encryption ensuring the relay operator cannot read token payloads; (3)~tier-aware context summarization that prevents long conversations from forcing simple queries onto expensive tiers; and (4)~an HPC-as-API proxy mode that exposes HPC inference as an OpenAI-compatible endpoint callable from any standard client with no HPC expertise, a deployment pattern made practical only by the sub-second TTFT of contribution~(2). Llama~3.2~3B achieves 85.1\% free-tier retention on a 1,200-query benchmark spanning ten domains. Measured TTFT: 0.26~s local, 0.54~s HPC (relay), 1.68~s cloud.
\end{abstract}



\maketitle

\begin{figure*}[t]
    \centering
    \includegraphics[width=\textwidth]{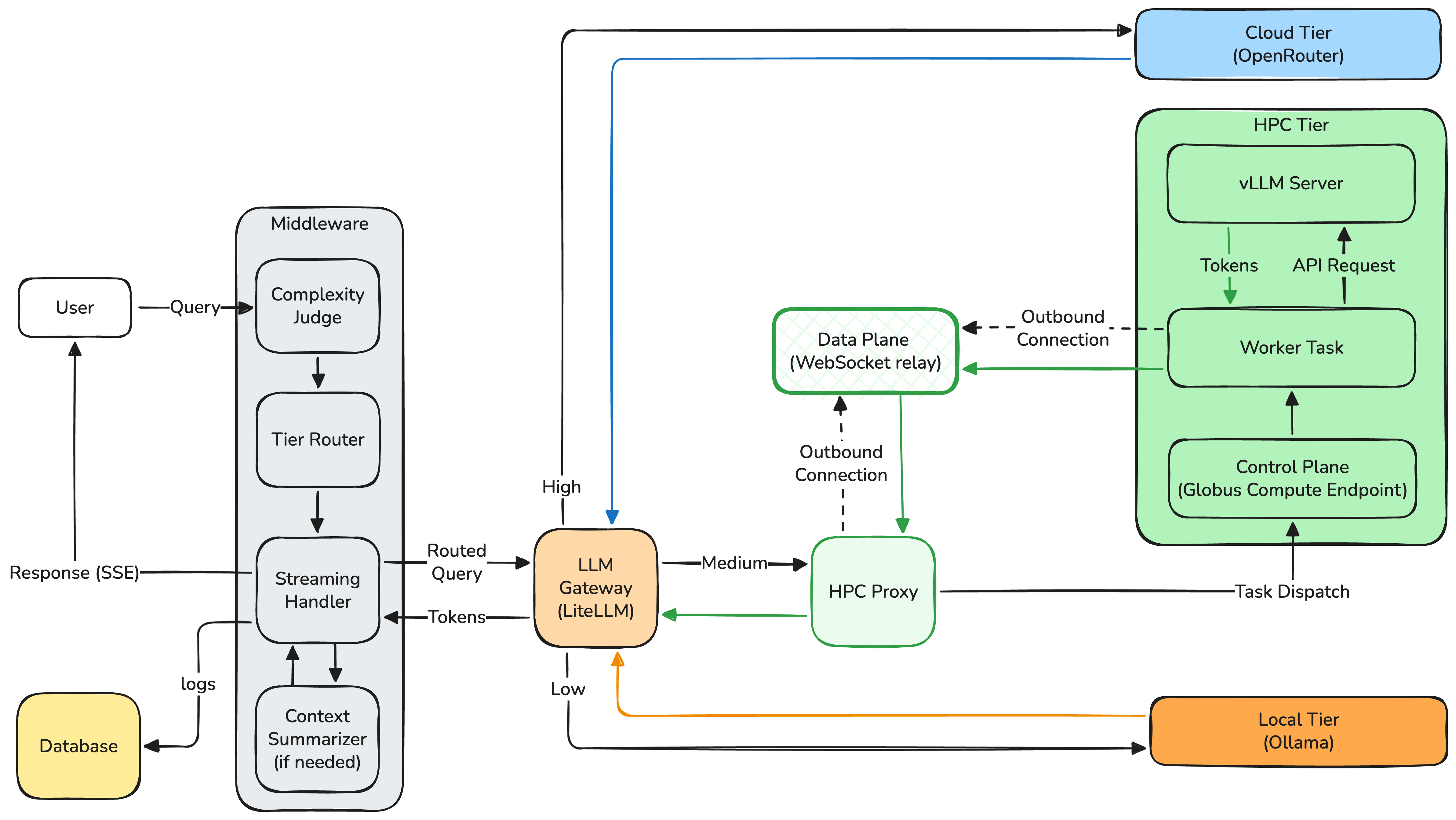}
    \caption{STREAM three-tier architecture (server mode). Queries are classified by the Complexity Judge and routed by the Tier Router to the appropriate tier: Local (Ollama), HPC, or Cloud. The Streaming Handler orchestrates the request, invoking the Context Summarizer if the conversation exceeds the target tier's context window, then forwards the query via the LiteLLM gateway and streams the response back to the user as SSE. For HPC queries, the HPC Proxy dispatches a task to the Control Plane (Globus Compute Endpoint), which executes a Worker Task that calls the vLLM Server over the cluster's internal network and connects \textit{outbound} to the Data Plane (WebSocket relay) as a token producer; the HPC Proxy independently connects \textit{outbound} to the relay as a consumer. Neither side accepts an inbound connection, traversing institutional firewalls without VPN or firewall rule changes.}
    \label{fig:architecture}
\end{figure*}


\section{Introduction}

Large language models are now central to research and education, yet students, researchers, and practitioners alike must choose between three imperfect options. Local inference is free and keeps data on-premises, but is constrained by the user's hardware: smaller models and limited context windows compared to what large GPU clusters can serve. Institutional HPC centers offer the GPU capacity to run large models at no marginal cost, but their firewall policies block standard streaming approaches, and their batch execution model provides no built-in mechanism for delivering incremental output to a waiting remote client. Commercial cloud APIs provide frontier-model quality on demand, but carry per-token costs and data retention policies that are unacceptable for sensitive research data. No single system spans all three, leaving researchers to maintain separate workflows and driving unnecessary cloud spending while institutional GPU resources sit underutilized. STREAM addresses this directly: by routing simple queries to free local models and only escalating to cloud when genuinely necessary, it reduces paid-tier usage while preserving access to frontier models for queries that need them. We present STREAM (Smart Tiered Routing Engine for AI Models), which unifies all three as a single tiered service and solves the HPC streaming problem: tokens are forwarded to the user in real time as the GPU generates them, reducing time-to-first-token (TTFT, the latency until the first output token reaches the user) from 11.40~s in batch mode (where TTFT equals the total response time) to 0.54~s with relay streaming, a 21.1$\times$ improvement. STREAM makes four contributions: (1)~\textbf{three-tier routing architecture}, the first system, to our knowledge, combining local device, institutional HPC, and cloud inference with automatic complexity-based routing (Section~\ref{sec:architecture}); (2)~\textbf{dual-channel HPC streaming}, a novel architecture that separates the Globus Compute control plane (authentication, dispatch) from a WebSocket relay data plane (real-time token delivery), enabling sub-second TTFT from HPC with end-to-end AES-256-GCM encryption (Section~\ref{sec:relay}); (3)~\textbf{tier-aware context summarization}, differential compression per tier preventing context-driven tier upgrades (Section~\ref{sec:compression}); and (4)~\textbf{HPC-as-API proxy mode}, which exposes HPC inference as an OpenAI-compatible endpoint callable from any standard client with no HPC expertise, a deployment pattern made practical only by the sub-second TTFT of contribution~(2) (Sections~\ref{sec:hpcasapi}--\ref{sec:security}).

\section{System Architecture}
\label{sec:architecture}

STREAM acts as a middleware layer between the user interface and three inference backends. Every query passes through the same pipeline: the complexity judge classifies it, the tier router selects the appropriate tier based on query complexity, and the streaming handler handles the rest: compressing older history if context nears the tier's limit, forwarding the query to the LiteLLM gateway, streaming the response token by token to the client, tracking token usage and cost, logging per-request metadata (model name, prompt token count, completion token count, and cost) to the database without storing any message content, and falling back automatically to the next tier on failure. Tokens are delivered over Server-Sent Events (SSE), a standard mechanism that lets a server push a continuous stream of text events over a single persistent HTTP connection. Figure~\ref{fig:architecture} shows the full architecture.

\subsection{Three Inference Tiers}

The \textbf{local} tier (Llama~3.2~3B / Gemma~3~4B-VL, 32K context, free) runs Ollama models on the user's device or a co-located server. The \textbf{HPC} tier (Qwen~2.5-VL-72B-AWQ, 64K context, free) runs as a persistent SLURM GPU job on an H100 NVL GPU (94~GB); a persistent Globus Compute~\cite{globuscompute2022} CPU worker stays alive on the cluster and forwards each query to the vLLM server over the cluster's internal network. vLLM~\cite{vllm2023} runs inside an Apptainer (formerly Singularity) container since HPC clusters do not permit Docker. The official vLLM~0.15.1 image targets CUDA~12.9, but the cluster's H100 NVL driver is CUDA~12.4 (driver 550); the resulting PTX version mismatch silently disables the Marlin AWQ kernels, cutting throughput to 20.1~tok/s. Building vLLM from source against CUDA~12.4 restores the Marlin kernels; combining this with prefix caching and chunked prefill yields the throughput reported in Table~\ref{tab:latency}. Finally, the \textbf{cloud} tier (300+ models via OpenRouter, 64K--1M context, usage cost) is the highest-capability option. All tiers share a unified OpenAI-compatible gateway (LiteLLM~\cite{litellm2023}).

\subsection{Complexity-Based Routing}

STREAM classifies each query as LOW, MEDIUM, or HIGH using a local \textbf{complexity judge} (Llama~3.2~3B), with a keyword fallback and result cache for repeated queries. LOW routes to local, MEDIUM to HPC, HIGH to cloud; users can override. Image queries substitute a vision-capable model (Gemma~3~4B-VL locally, Qwen~2.5-VL-72B on HPC) without changing the routing decision.

Health checking avoids a latency trap: STREAM performs only a lightweight Globus authentication check ($\approx$100~ms) at routing time. If HPC turns out to be unreachable when streaming starts, the handler falls back to the next tier automatically. Fallback chains are asymmetric: MEDIUM queries escalate HPC~$\to$~cloud~$\to$~local; HIGH queries descend cloud~$\to$~HPC~$\to$~local, preserving quality expectations per class.

\subsection{Deployment Modes}

STREAM ships in two deployment modes that share over 90\% of the codebase. \textbf{Server mode} runs five Docker containers (middleware, LLM gateway, Ollama, PostgreSQL, and the HPC proxy), suited for shared deployments. \textbf{Desktop mode} collapses everything into a single PyWebView process backed by SQLite with no external dependencies. A third standalone deployment, \textbf{HPC-as-API mode}, exposes the HPC tier as an independent public endpoint, described in Section~\ref{sec:hpcasapi}.


\section{Dual-Channel HPC Streaming}
\label{sec:relay}

HPC job schedulers are designed for batch workloads: a client submits a job, waits for it to finish, and retrieves the complete result. There is no built-in mechanism for streaming partial output during execution. STREAM uses Globus Compute~\cite{globuscompute2022} as its control plane, which handles federated authentication and remote function execution across HPC resources, but it inherits this same batch model. Globus also offers a separate Globus Streams service; comparing the two is left to future work.

\textbf{The networking challenge.} Streaming tokens from HPC is not just a software problem; it is a networking one. HPC compute nodes sit behind institutional firewalls that block all inbound connections. Client machines are typically behind Network Address Translation (NAT), where a router assigns many internal addresses to one shared external IP, making them unreachable from outside without explicit firewall configuration. While a dedicated cloud VM can accept inbound connections, registering its address and opening a port for every new job and user does not scale across institutions. Standard streaming approaches require at least one inbound-capable endpoint and therefore fail in this environment without per-institution VPN or firewall changes.

\subsection{Control Plane / Data Plane Separation}

STREAM overcomes the firewall constraint by splitting each HPC request across two independent planes:

\begin{enumerate}
  \item \textbf{Control plane} (Globus Compute): STREAM authenticates the user via Globus Auth's OAuth2 federation and submits a task containing the conversation messages, target model, inference parameters, the vLLM and relay URLs, and a freshly generated UUID channel ID. No relay credentials travel through the control plane: both the relay shared secret (channel access token) and the AES-256-GCM payload encryption key (confidentiality credential) are pre-loaded into the remote process environment and never passed as task arguments, so neither ever appears in any task record. When the relay is reachable, Globus Compute carries the query to the HPC cluster but delivers no token output back; that is the data plane's job. Tokens traverse the control plane only in the batch fallback path, when the relay is unavailable (Section~\ref{sec:evaluation}).

  \item \textbf{Data plane} (WebSocket relay): For every query, the proxy generates a fresh UUID channel ID, submits the Globus Compute task with that ID as an argument, then immediately opens a \textit{consumer} connection to that channel on the relay. Concurrently, once the task begins executing on the CPU node, it connects outbound as a \textit{producer} and reads the vLLM HTTP SSE stream token by token over the cluster's internal network, forwarding each token to the relay in real time. The channel ID is the shared rendezvous: both sides connect to the same UUID without any direct knowledge of each other's address. Because Globus Compute takes a few hundred milliseconds to dispatch the task, the locally-opened consumer almost always connects first and is waiting when the first token arrives. If the producer is faster, the relay buffers up to 1,000 tokens per channel and replays them in order the moment the consumer connects, so no tokens are dropped. In server mode the consumer is the dedicated HPC Proxy container (Figure~\ref{fig:architecture}); in desktop mode it is \texttt{litellm\_direct} in the same process as the middleware.
\end{enumerate}

The key insight is that \textbf{both producer and consumer connect outbound}; neither side ever accepts an inbound connection. This is what makes the architecture firewall-transparent: HPC compute nodes can initiate outbound connections to the public internet, but institutional firewalls block all inbound connections to those nodes. By having both endpoints reach out to the relay — a cloud VM with a stable public address — STREAM avoids the need for any firewall rule changes, open ports, or VPN configuration on the HPC side. Outbound \texttt{wss://} on port~443 is universally permitted by institutional firewall policy. Channel IDs are UUIDs (122 bits of entropy), making unauthorized channel guessing computationally infeasible even without authentication. If the relay is unavailable, STREAM falls back to batch mode: Globus Compute returns the complete response as a single payload and TTFT equals total response time (Section~\ref{sec:evaluation}).

\subsection{Relay Protocol and Implementation Notes}

Channels are \textbf{per-query and stateless}: a fresh UUID channel is created for every query, both sides disconnect at completion and the channel is removed immediately; if one side never connects (e.g., a failed Globus Compute task), the channel is reaped after a 300~s timeout sized to cover the worst-case cold-start delay. Per-query channels are the natural fit: one query, one stream, one channel lifetime. Sharing a channel across queries would require the relay to track sessions and sequence multiple streams; it becomes a stateful broker rather than a simple forwarding pipe. Two non-obvious integration issues arose during implementation: (1)~\textbf{serialization:} Globus Compute serializes functions with \texttt{dill}; in desktop mode, PyInstaller-packaged builds reference an internal import system that \texttt{dill} cannot resolve on the endpoint. Fix: ship the function as a source string executed with \texttt{exec()}. Server mode reuses this path to avoid code duplication. (2)~\textbf{remote package availability:} STREAM is not installed on the HPC endpoint, so its AES-256-GCM encryption helper cannot be imported there; the helper is copied directly into the remote function body instead.

\section{HPC-as-API Proxy Mode}
\label{sec:hpcasapi}

The dual-channel architecture gives STREAM users sub-second HPC streaming, but external callers such as a cloud-hosted application or a desktop tool cannot be expected to obtain Globus credentials or implement the relay WebSocket protocol directly. STREAM's HPC proxy removes this barrier: it accepts a standard \texttt{POST /v1/chat/completions} request, runs the full dual-channel flow internally, and returns an OpenAI-compatible SSE stream. Callers need only a bearer token and a base URL.

\textbf{Dual-mode authentication.} The proxy authenticates callers through a single \texttt{Authorization: Bearer} header supporting two coexisting modes. \textit{Globus Token Auth} is for institutional users: the proxy verifies the caller's Globus access token with the Globus Auth service, confirms the email domain (e.g., \texttt{@uic.edu}), and submits the Globus Compute job under the caller's own Globus identity without per-user enrollment on the proxy. \textit{API Key Auth}, by contrast, is for external services (e.g., a cloud-hosted application that manages its own user authentication): the proxy validates a pre-issued key and submits jobs under its own Globus Compute credentials stored on the proxy. The proxy tries Globus token verification first and falls back to API key lookup automatically. In both modes, every request is logged with the caller identity, credential hash, and client IP; a per-caller sliding-window rate limit and message format validation (role names, content length, message count) are enforced before any job reaches the cluster.

\textbf{Deployment scope and limitations.} The proxy and relay share one VM; production deployments with availability SLAs should run them separately. The current design suits a small set of known teams with manually issued keys. The novelty is not the proxy pattern itself but that \textbf{it is viable}: without streaming, a proxy over a batch HPC call imposes 11+ seconds of latency, making it unusable for interactive applications. The 0.54~s TTFT is what transforms this from a theoretical pattern into a practical deployment.


\section{Security}
\label{sec:security}

STREAM's security design treats the relay VM as an untrusted intermediary and builds three independent layers such that compromising any one layer does not expose user data.

\textbf{Transport security (TLS).} All connections operate over \texttt{wss://} or \texttt{https://}, with TLS termination via Caddy (an open-source reverse proxy) using Let's Encrypt (a free, automated certificate authority) for certificate management. No plaintext traffic crosses any network boundary.

\textbf{Access control.} The proxy and relay enforce independent access control. The proxy validates a bearer token on every incoming request and rejects it before any Globus Compute work is initiated, so unauthenticated callers cannot trigger HPC jobs. The relay validates a pre-shared secret on every producer and consumer connection. Critically, the secret is transmitted as the \textit{first JSON message after the WebSocket handshake completes}, not as a URL query parameter. This is a non-obvious but important detail: URL query parameters (e.g., \texttt{?secret=...}) appear in server access logs because web servers log the full request URL after TLS decryption; \texttt{?secret=} would be written to disk in plaintext on the relay VM. Post-handshake transmission keeps the secret out of all log files. Connections that do not supply a valid auth message within 10~seconds are closed. No Globus Auth credentials traverse the relay at any point.

\textbf{End-to-end payload encryption (AES-256-GCM).} TLS protects the link to the relay but leaves payloads visible to the relay operator. STREAM optionally encrypts every token payload: the producer generates a fresh 12-byte random nonce per message, encrypts with AES-256-GCM, appends a 16-byte authentication tag, and sends the base64-encoded ciphertext in a JSON envelope. The relay forwards opaque ciphertext; a tampered payload is detected at the consumer before reaching the user. Both the relay shared secret (\texttt{RELAY\_SECRET}, channel-access token) and the AES-256-GCM key (\texttt{RELAY\_ENCRYPTION\_KEY}, confidentiality credential) are pre-provisioned in the Globus Compute endpoint's \texttt{worker\_init} environment and read from \texttt{os.environ} inside the remote function; neither is ever a task argument, so neither traverses the relay or AMQP. Globus credentials never leave the proxy VM.

Together these layers address the three principal threat surfaces: the network path (TLS), unauthorized access (access control), and a compromised relay operator (AES-256-GCM).

\section{Tier-Aware Context Summarization}
\label{sec:compression}

Existing LLM routing systems treat each query as an independent decision, ignoring the conversation history that has built up before it~\cite{routellm2024,hybridllm2024,frugalgpt2023}. This creates a subtle but costly problem: as a conversation grows longer, even a trivially simple query like ``What is 2+2?'' may exceed the local tier's context window. The system is then forced to route it to a more expensive tier, not because the query is complex, but simply because the history is long.

STREAM addresses this with \textit{tier-aware rolling summarization}. When a conversation approaches 80\% of the target tier's context limit, the system compresses older messages into a summary while keeping recent turns verbatim. The compression settings are calibrated to each tier's context window: the \textbf{local} tier (32K tokens) uses a 2K-token summary plus the last 3 turn pairs; the \textbf{HPC} tier (64K) uses a 4K-token summary plus the last 6 turn pairs; the \textbf{cloud} tier disables summarization by default since its windows are large enough to rarely need it. Summarization is performed by the free local model (Llama~3.2~3B) at zero additional cost.


\section{Evaluation}
\label{sec:evaluation}

\textbf{Methodology.} Latency benchmarks report medians over 50 single-turn requests using \textit{tier bypass mode} (routing judge disabled), isolating per-tier latency. TTFT (time to first token — the latency until the first token reaches the client) is measured from request submission to first SSE event. Local benchmarks ran on an Apple M4 Pro (24~GB unified memory); HPC benchmarks used a dedicated single-user Globus Compute endpoint (no queue contention). Routing evaluation uses a balanced 1,200-query held-out test set (400 per complexity class: LOW, MEDIUM, HIGH) drawn from real user questions in publicly available research and community datasets spanning ten domains.\footnote{Domains: HPC, mathematics, statistics/ML, physics/chemistry, engineering, life sciences, CS/software, philosophy/ethics, social sciences, history/culture. Sources: StackExchange~\cite{stackexchange}, MMLU~\cite{mmlu2021}, MMLU-Pro~\cite{mmlupro2024}, PubMedQA~\cite{pubmedqa2019}.} Since STREAM has no real-user deployment traces yet, ground-truth complexity labels were assigned by Claude Sonnet~4.6; the runtime complexity judge (Llama~3.2~3B) is then evaluated against those labels. Complexity is defined by \textit{reasoning depth} (LOW: single retrievable answer; MEDIUM: 2--4 concepts assembled; HIGH: novel reasoning path or expert judgment); the difficulty of solving the query, not its length.\footnote{\url{https://huggingface.co/datasets/anasnassar/llm-query-complexity-benchmark}}

\subsection{Routing Accuracy}

The key cost metric is \textbf{paid-tier leakage}: LOW or MEDIUM queries misclassified as HIGH and routed to the paid cloud tier. LOW-to-MEDIUM misroutes are financially harmless since both tiers are free. A random classifier achieves 33.3\% on this three-class problem.

Table~\ref{tab:routing} shows the confusion matrix for Llama~3.2~3B. The judge achieves 49.0\% overall accuracy, 1.5$\times$ above chance, with 119 leaked queries and \textbf{85.1\% free-tier retention}. Its main weakness is LOW recall (8.2\%): nearly all simple queries are over-routed to MEDIUM. This is a conservative error since both local and HPC are free, but it does mean simple queries run on a larger model than necessary. The judge adds 164~ms median latency per query (p95=230~ms), paid once in AUTO mode before any tier dispatch. Overall, 49\% accuracy leaves significant room for improvement; replacing the LLM-as-a-judge approach with a dedicated fine-tuned text classifier is the most important next step for STREAM's routing quality.

\begin{table}[t]
\centering
\caption{Llama~3.2~3B routing confusion matrix (164~ms median judge latency, 49.0\% overall, 85.1\% free-tier retention, 119 leaked queries).}
\label{tab:routing}
\small
\begin{tabular}{@{}lrrrr@{}}
\toprule
\textbf{True$\backslash$Pred} & \textbf{LOW} & \textbf{MED} & \textbf{HIGH} & \textbf{Recall} \\
\midrule
LOW    &  33 & 328 &  39 &  8.2\% \\
MEDIUM &   6 & 314 &  80 & 78.5\% \\
HIGH   &   1 & 158 & 241 & 60.2\% \\
\midrule
Precision & 82.5\% & 39.3\% & 66.9\% & F1: 0.44 \\
\midrule
\multicolumn{4}{l}{Overall (588/1200)} & 49.0\% \\
\bottomrule
\end{tabular}
\end{table}

\subsection{Response Latency}

Table~\ref{tab:latency} reports TTFT and throughput for each tier. The local tier achieves 0.26~s median TTFT at 85.5~tok/s. The HPC tier achieves 0.54~s median TTFT via relay streaming ($\pm$0.34~s, p95=1.53~s), sub-second despite two network hops and a Globus Compute dispatch, because the relay delivers the first token the moment vLLM generates it. In batch fallback mode (relay disabled), TTFT equals total generation time: 11.40~s median ($\pm$3.27~s, p95=15.87~s) across 50 diverse queries with a warm server (vLLM running, weights resident in GPU VRAM). Both modes achieve the same 26.9~tok/s generation throughput (Table~\ref{tab:latency}), confirming the relay adds no per-token overhead. The cloud tier (Claude Sonnet~4.6 via OpenRouter) reaches 1.68~s median TTFT ($\pm$0.52~s).

\begin{table}[t]
\centering
\caption{Response latency (medians over 50 runs, complexity judge bypassed). $\dagger$Batch fallback: relay disabled; TTFT equals total generation time (response-length dependent). HPC: Qwen~2.5-VL-72B-AWQ on H100 NVL 94~GB, warm server. Cloud: Claude Sonnet~4.6 via OpenRouter.}
\label{tab:latency}
\small
\begin{tabular}{@{}lrr@{}}
\toprule
\textbf{Tier} & \textbf{TTFT (s)} & \textbf{tok/s} \\
\midrule
Local (Llama 3.2 3B)            & $0.26 \pm 0.07$  & 85.5 \\
HPC (relay streaming)           & $0.54 \pm 0.34$  & 26.9 \\
HPC (batch, warm server)$\dagger$      & $11.40 \pm 3.27$  & 26.9 \\
Cloud (Claude Sonnet~4.6)       & $1.68 \pm 0.52$  & 41.8 \\
\bottomrule
\end{tabular}
\end{table}

\subsection{Context Summarization}

We pre-filled five 40-turn conversations ($\sim$1,050 tokens/turn) and sent the probe \textit{``What is 2+2?''} at turns 10--40 with and without summarization. Without summarization, the conversation history exceeds the local tier's 32K context limit at turn~30, and every subsequent trivial query is forced to the HPC tier. With summarization enabled, STREAM compresses the older history to $\approx$4,300 tokens (a 90\% reduction) the moment context reaches 80\% of the tier limit; the probe stayed on the local tier at every turn tested, including turn~40 where the raw context was 41K tokens.

\begin{table}[h]
\centering
\caption{Context summarization results. Five 40-turn synthetic conversations ($\sim$1,050 tokens/turn). Probe: ``What is 2+2?'' (LOW complexity). $\dagger$Exceeds local tier's 32K context limit.}
\label{tab:compression}
\small
\begin{tabular}{@{}lcrr@{}}
\toprule
\textbf{Turn} & \textbf{$\approx$Tokens} & \textbf{No Summ.} & \textbf{With Summ.} \\
\midrule
10  & 10.5K & Local   & Local \\
20  & 21K   & Local   & Local \\
30  & 31.6K & Upgraded$\dagger$ & Local \\
35  & 36.3K & Upgraded$\dagger$ & Local \\
40  & 41K   & Upgraded$\dagger$ & Local \\
\midrule
\multicolumn{2}{l}{First forced upgrade} & Turn 30 & \textbf{Never} \\
\bottomrule
\end{tabular}
\end{table}

\subsection*{Limitations}
STREAM is a research prototype; the evaluations establish a performance baseline and identify directions for improvement, not production-grade characterizations. Routing accuracy reflects a benchmark labeled by a single LLM annotator (Claude Sonnet~4.6); labels from a single model without human validation or inter-annotator agreement measurement introduce potential bias in the ground truth. Latency was measured on a dedicated single-user endpoint; shared deployments with concurrent users may see higher TTFT due to Globus Compute worker queuing. Summarization was evaluated on synthetic conversations, which may not reflect real usage patterns.


\section{Related Work}

FIRST~\cite{first2025} is the closest prior work, dispatching LLM inference to HPC via Globus Compute with federated authentication, but does not report TTFT or describe how tokens are streamed through Globus Compute's task-oriented model. STREAM's dual-channel relay addresses this gap. Chat~AI~\cite{chatai2024} connects a chat interface to Slurm-scheduled HPC using SSH-based access control, without a federated identity layer. Institutional deployments at Dartmouth~\cite{dartmouth2025} and Purdue~\cite{purdue2025} demonstrate demand for on-premises LLM access but implement only single-tier inference without complexity-based routing across tiers.

In the LLM routing space, RouteLLM~\cite{routellm2024}, Hybrid LLM~\cite{hybridllm2024}, and FrugalGPT~\cite{frugalgpt2023} each learn routing policies between strong and weak models over commercial API endpoints, treating routing as a stateless per-query cost-quality tradeoff. None integrate HPC resources or account for context window differences across tiers.

Existing LLM memory systems manage long-term memory without adapting to different tiers' context limits; STREAM's tier-aware summarization is, to our knowledge, the first to tailor compression depth per tier to prevent context-driven tier upgrades.


\section{Conclusion}
\label{sec:conclusion}

STREAM demonstrates that local, HPC, and cloud LLM inference can be unified in a single system with interactive response times across all three tiers.\footnote{Code: \url{https://github.com/uicacer/STREAM}. Demo: \url{https://www.youtube.com/watch?v=eTllhqNtOOY}.} The dual-channel relay brings HPC TTFT from 11.40~s down to 0.54~s (21.1$\times$), end-to-end AES-256-GCM encryption keeps token payloads opaque to the relay, and tier-aware summarization sustains local-tier routing through 40+ turns. The relay and proxy are released as domain-agnostic standalone PyPI libraries — \texttt{streamrelay}~\cite{nassar2026streamrelay} and \texttt{hpc-as-api}~\cite{nassar2026hpcasapi} — archived on Zenodo, generalizing to any HPC job that produces incremental output. By abstracting away HPC job submission, firewall traversal, and credential management, STREAM's HPC-as-API proxy mode makes institutional GPU resources as accessible as a cloud API — lowering the barrier for researchers who need large-model inference without the cost. Future work includes replacing the LLM-as-a-judge complexity judge with a fine-tuned text classifier, collecting real-user routing traces, a direct comparison against alternative HPC streaming approaches (Globus Streams, SciStream), and per-user SLURM attribution via a Multi-User Endpoint.

\begin{acks}
The authors thank Marius Horga and Lanre Adio at ACER (UIC) for support and relay server infrastructure. This research used resources of the Lakeshore HPC cluster at the University of Illinois Chicago.
\end{acks}

\subsection*{AI Tool Disclosure}
Claude Code (Anthropic) was used for code generation, debugging, and paper drafting. All architectural decisions, system design, and evaluation methodology are the authors' original work.


\bibliographystyle{ACM-Reference-Format}
\bibliography{references}

@inproceedings{first2025,
  author    = {Aditya Tanikanti and Benoit C{\^o}t{\'e} and Yanfei Guo and Le Chen and Nickolaus Saint and Ryan Chard and Ken Raffenetti and Rajeev Thakur and Thomas Uram and Ian Foster and Michael E. Papka and Venkatram Vishwanath},
  title     = {{FIRST}: Federated Inference Resource Scheduling Toolkit for Scientific {AI} Model Access},
  booktitle = {Proceedings of the SC '25 Workshops},
  year      = {2025},
  publisher = {ACM},
  pages     = {52--60},
  doi       = {10.1145/3731599.3767346},
}

@article{chatai2024,
  author  = {Ali Doosthosseini and Jonathan Decker and Hendrik Nolte and Julian M. Kunkel},
  title   = {Chat {AI}: A Seamless Slurm-Native Solution for {HPC}-Based Services},
  journal = {arXiv preprint arXiv:2407.00110},
  year    = {2024},
  doi     = {10.48550/arXiv.2407.00110},
}

@inproceedings{dartmouth2025,
  author    = {Simon Stone and Jonathan Crossett and Tivon Luker and Lora Leligdon and William Cowen and Christian Darabos},
  title     = {Dartmouth Chat - Deploying an Open-Source {LLM} Stack at Scale},
  booktitle = {Proceedings of the Practice and Experience in Advanced Research Computing Conference (PEARC '25)},
  year      = {2025},
  publisher = {ACM},
  pages     = {1--5},
  doi       = {10.1145/3708035.3736076},
}

@inproceedings{purdue2025,
  author    = {Sarah Rodenbeck and Erik Gough and Athreyan Mohana Krishnan Sangeetha and {Ashish} and Mihir Ahlawat and Vivek Karunai Kiri Ragavan and Abhishek Muthukumar and Aanis Ahmad},
  title     = {Providing On-Prem {GenAI} Inference Services to a Campus Community},
  booktitle = {Proceedings of the Practice and Experience in Advanced Research Computing Conference (PEARC '25)},
  year      = {2025},
  publisher = {ACM},
  pages     = {1--4},
  doi       = {10.1145/3708035.3736039},
}

@article{routellm2024,
  author  = {Isaac Ong and Amjad Almahairi and Vincent Wu and Wei-Lin Chiang and Tianhao Wu and Joseph E. Gonzalez and M Waleed Kadous and Ion Stoica},
  title   = {{RouteLLM}: Learning to Route {LLMs} with Preference Data},
  journal = {arXiv preprint arXiv:2406.18665},
  year    = {2024},
}

@inproceedings{hybridllm2024,
  author    = {Dujian Ding and Ankur Mallick and Chi Wang and Robert Sim and Subhabrata Mukherjee and Victor Ruhle and Laks V. S. Lakshmanan and Ahmed Hassan Awadallah},
  title     = {Hybrid {LLM}: Cost-Efficient and Quality-Aware Query Routing},
  booktitle = {Proceedings of the International Conference on Learning Representations (ICLR)},
  year      = {2024},
}

@article{globuscompute2022,
  author  = {Zhuozhao Li and Ryan Chard and Yadu Babuji and Ben Galewsky and Tyler J. Skluzacek and Kirill Nagaitsev and Anna Woodard and Ben Blaiszik and Josh Bryan and Daniel S. Katz and Ian Foster and Kyle Chard},
  title   = {{funcX}: Federated Function as a Service for Science},
  journal = {IEEE Transactions on Parallel and Distributed Systems},
  volume  = {33},
  number  = {12},
  pages   = {4948--4963},
  year    = {2022},
  doi     = {10.1109/TPDS.2022.3208767},
}

@article{frugalgpt2023,
  author  = {Lingjiao Chen and Matei Zaharia and James Zou},
  title   = {{FrugalGPT}: How to Use Large Language Models While Reducing Cost and Improving Performance},
  journal = {Transactions on Machine Learning Research},
  year    = {2024},
  url     = {https://openreview.net/forum?id=cSimKw5p6R},
}

@misc{litellm2023,
  author       = {{BerriAI}},
  title        = {{LiteLLM}: Call 100+ {LLM} {APIs} in {OpenAI} Format},
  year         = {2023},
  url          = {https://github.com/BerriAI/litellm},
}

@inproceedings{vllm2023,
  author    = {Woosuk Kwon and Zhuohan Li and Siyuan Zhuang and Ying Sheng and Lianmin Zheng and Cody Hao Yu and Joseph E. Gonzalez and Hao Zhang and Ion Stoica},
  title     = {Efficient Memory Management for Large Language Model Serving with {PagedAttention}},
  booktitle = {Proceedings of the ACM SIGOPS 29th Symposium on Operating Systems Principles (SOSP '23)},
  year      = {2023},
  publisher = {ACM},
  pages     = {611--626},
  doi       = {10.1145/3600006.3613165},
}

@software{nassar2026streamrelay,
  author    = {Anas Nassar},
  title     = {{streamrelay}: {WebSocket} Relay for Real-Time Streaming from Batch {HPC} Executors},
  year      = {2026},
  publisher = {Zenodo},
  version   = {0.3.0},
  doi       = {10.5281/zenodo.20542696},
  url       = {https://doi.org/10.5281/zenodo.20542696},
}

@software{nassar2026hpcasapi,
  author    = {Anas Nassar},
  title     = {{hpc-as-api}: An {OpenAI}-Compatible {API} Gateway for {HPC} Clusters via {Globus Compute}},
  year      = {2026},
  publisher = {Zenodo},
  version   = {0.3.2},
  doi       = {10.5281/zenodo.20545439},
  url       = {https://doi.org/10.5281/zenodo.20545439},
}

@misc{stackexchange,
  author = {{Stack Exchange, Inc.}},
  title  = {Stack Exchange Data Dump},
  year   = {2024},
  url    = {https://archive.org/details/stackexchange},
}

@inproceedings{mmlu2021,
  author    = {Dan Hendrycks and Collin Burns and Steven Basart and Andy Zou and Mantas Mazeika and Dawn Song and Jacob Steinhardt},
  title     = {Measuring Massive Multitask Language Understanding},
  booktitle = {Proceedings of the 9th International Conference on Learning Representations (ICLR 2021)},
  year      = {2021},
  url       = {https://openreview.net/forum?id=d7KBjmI3GmQ},
}

@article{mmlupro2024,
  author  = {Yubo Wang and Xueguang Ma and Ge Zhang and others},
  title   = {{MMLU-Pro}: A More Robust and Challenging Multi-Task Language Understanding Benchmark},
  journal = {arXiv preprint arXiv:2406.01574},
  year    = {2024},
  url     = {https://arxiv.org/abs/2406.01574},
}

@inproceedings{pubmedqa2019,
  author    = {Qiao Jin and Bhuwan Dhingra and Zhengping Liu and William W. Cohen and Xinghua Lu},
  title     = {{PubMedQA}: A Dataset for Biomedical Research Question Answering},
  booktitle = {Proceedings of the 2019 Conference on Empirical Methods in Natural Language Processing and the 9th International Joint Conference on Natural Language Processing (EMNLP-IJCNLP)},
  pages     = {2567--2577},
  year      = {2019},
  publisher = {Association for Computational Linguistics},
  doi       = {10.18653/v1/D19-1259},
  url       = {https://aclanthology.org/D19-1259/},
}

\end{document}